\documentclass[usenatbib]{mn2e}
\usepackage{graphicx, amssymb, aas_macros, natbib, bm}
\usepackage{ marvosym }
\usepackage{enumitem}

\newcommand{\mstar}{{M}_{\star}}

\newcommand{\msun}{{\rm M}_{\odot}}

\newcommand{\gt}{>}

\title[Ch-Ch-Changing Satellites via Stripping]
{Under Pressure: Quenching Star Formation in Low-Mass Satellite Galaxies via
  Stripping}

\author[Fillingham et al.]
{Sean P. Fillingham,$^1$\thanks{$\!\!$e-mail: sfilling@uci.edu}
Michael C. Cooper,$^1$\thanks{$\!\!$e-mail: cooper@uci.edu} 
Andrew B. Pace,$^1$
\newauthor
Michael Boylan-Kolchin,$^2$
James S. Bullock,$^1$
Shea Garrison-Kimmel,$^3$\thanks{$\!\!$Einstein fellow} 
\newauthor 
Coral Wheeler$^1$ \\
$\!\!^1$Center for Cosmology, Department of Physics \& Astronomy,
4129 Reines Hall, University of California, Irvine, CA 92697, USA \\
$\!\!^2$Department of Astronomy, The University of Texas at Austin,
2515 Speedway, Stop C1400, Austin, TX 78712, USA \\
$\!\!^3$TAPIR, Mailcode 350-17, California Institute of Technology,
Pasadena, CA 91125, USA} 

\begin{document}

\pagerange{\pageref{firstpage}--\pageref{lastpage}} 
\pubyear{2016}

\maketitle

\label{firstpage}
\begin{abstract}

  Recent studies of galaxies in the local Universe, including those in
  the Local Group, find that the efficiency of environmental (or
  satellite) quenching increases dramatically at satellite stellar
  masses below $\sim10^{8}~\msun$. This suggest a physical scale
  where quenching transitions from a slow ``starvation'' mode to a
  rapid ``stripping'' mode at low masses.
  We investigate the plausibility of this scenario using observed
  H{\scriptsize I} surface density profiles for a sample of $66$
  nearby galaxies as inputs to analytic calculations of ram-pressure
  and turbulent viscous stripping. Across a broad range of host properties, we
  find that stripping becomes increasingly effective at
  $M_{*}\lesssim10^{8-9}~\msun$, reproducing the critical mass scale
  observed.
  However, for canonical values of the circumgalactic medium density
  ($n_{\rm halo} < 10^{-3.5}$~cm$^{-3}$), we find that stripping is
  not fully effective; infalling satellites are, on average, stripped
  of only $\lesssim40-60\%$ of their cold gas reservoir, which is
  insufficient to match observations.
  By including a host halo gas distribution that is clumpy and
  therefore contains regions of higher density, we are able to
  reproduce the observed H{\scriptsize I} gas fractions (and thus the
  high quenched fraction and short quenching timescale) of Local Group
  satellites, suggesting that a host halo with clumpy gas may be
  crucial for quenching low-mass systems in Local Group-like (and more
  massive) host halos.

\end{abstract}


\section{Introduction}
\label{sec:intro} 

Recent studies probing the properties of satellite galaxies in the
local Universe show that the suppression (or ``quenching'') of star
formation in satellites is a relatively inefficient process relative
to the expectations of hydrodynamic and semi-analytic models of galaxy
formation \citep[e.g.][]{kimm09, kimm11, wang14, hirschmann14,
  phillips15a}.
While satellites are rapidly quenched -- following infall -- in the
models, analysis of satellite populations identified in the Sloan
Digital Sky Survey \citep[SDSS,][]{york00} instead find that quenching
proceeds remarkably slowly, such that a typical satellite with $\mstar
\gtrsim 10^{8}~\msun$ orbits within its host halo -- continuing to
form stars -- for $\sim3-7$~Gyr before being quenched
\citep{delucia12, wetzel13, wheeler14}.\footnote{These measured
  satellite quenching timescales include the transition of a satellite
  system from star-forming to quiescent, which must proceed quickly
  (within $\sim1$~Gyr) so as to reproduce the observed bimodal
  distribution of specific star-formation rates and rest-frame colors
  \citep{balogh04, wetzel13}.}
Only at the lowest satellite masses is quenching a highly efficient
process, with low-mass ($\mstar < 10^{8}~\msun$) satellites in the
Local Group quenching within $\sim1-2$~Gyr of infall \citep{weisz15,
  wetzel15b, fham15}.

In an effort to connect these measured quenching timescales to the
relevant physical mechanisms at play, \citet{fham15} present
a comprehensive picture of satellite quenching spanning
roughly $5$ orders of magnitude in satellite stellar mass.
The low efficiency and long quenching timescales inferred for
intermediate- and high-mass satellites ($\mstar \gtrsim 10^{8}~\msun$)
are consistent with quenching via starvation -- a scenario in which
gas accretion on to a satellite galaxy is halted following infall,
thus eventually eliminating the fuel for star formation
\citep{larson80, kawata08};
lending support to this picture, the measured quenching timescales
agree very well with the observed cold gas (H{\scriptsize I} +
H$_{2}$) depletion timescales for field systems at $z \sim 0$
\citep{fham15}.
At lower satellite masses ($\mstar \lesssim 10^{8}~\msun$), however,
the quenching timescales derived from analysis of the Local Group
satellite population suggest that the physics of satellite quenching
must change significantly; a more efficient quenching mechanism
(relative to starvation) must be at play below a critical mass scale
of $\sim10^{8}~\msun$.

Stripping is a plausible candidate quenching mechanism at low
masses. This includes ram-pressure stripping \citep{gunn72}, a
process by which the cool, dense interstellar medium (and thus the
fuel for future star formation) is removed from a satellite galaxy as
it passes through its host's circumgalactic medium (CGM).
Ram-pressure stripping becomes increasingly effective in lower-mass
satellites, due to their weaker gravitational restoring pressures
\citep{hester06}; moreover, ram pressure acts on roughly the dynamical
time of the host system \citep[i.e.~$1-3$~Gyr,][]{tonnesen07,
  bekki14}, consistent with the short quenching timescale inferred for
low-mass satellites of the Local Group.
In addition to ram-pressure stripping, cold gas may also be removed
from a satellite due to turbulent viscous stripping, which results
from Kelvin-Helmholtz instabilities at the interface of the
satellite's interstellar medium and the CGM \citep{nulsen82}.
The relative motion of the two media in addition to the substantial
difference in their mean densities can lead to perturbations that
overcome the local gravitational restoring force, such that gas is
stripped from the satellite.
Within massive groups and clusters, a wide range of observations
provide abundant evidence of stripping in action, showing its ability
to quench infalling galaxies via removal of their cold gas component
\citep[e.g.][]{ebeling14, kenney15}.
It remains uncertain, however, if stripping is an
effective quenching mechanism in more typical host halos, such as that
of the Milky Way or M31, and specifically at satellite stellar masses
of $\lesssim10^{8}~\msun$.

In this work, our goal is to directly address the efficacy of
stripping as a quenching mechanism for low-mass satellites in Milky
Way-like systems. By using observations of local field dwarfs to
inform analytic calculations of both ram-pressure and turbulent
viscous stripping, we measure the amount of cold gas that would
typically be removed \emph{if} these field dwarfs were to interact
with a Milky Way-like host.
In Section~\ref{sec:rps}, we detail our methods, including the
analytic framework and data sets that we utilize to estimate the
impact of stripping on infalling satellites.
In Section~\ref{sec:results}, we present our primary results regarding
the efficiency of stripping in Milky Way-like environments,
specifically addressing potential uncertainties associated with the
properties of the host halo and the satellite population.
Finally, in Sections~\ref{sec:disc} and \ref{sec:endgame}, we discuss
and summarize the implications of our results with regard to the
quenching of low-mass satellites in the Local Group and beyond.

\section{Testing Satellite Stripping}
\label{sec:rps}

\subsection{Analytic Framework}
\label{subsec:af}
The effectiveness of stripping as a quenching mechanism boils down to
a relatively simple competition between the stripping pressure
($P_{\rm stripping}$) and the gravitational restoring force per unit
area (i.e.~the gravitational restoring pressure, $P_{\rm restore}$).
When $P_{\rm restore} \ge P_{\rm stripping}$, the interstellar medium
(or ISM, comprised predominantly of cold gas) is retained by the
infalling satellite galaxy, such that star formation may proceed.
When the stripping pressure exceeds the gravitational restoring
pressure, however, some fraction of the cold gas is removed from the
satellite. In cases where this stripped fraction is large enough, star
formation will be shut down rapidly due to the loss of available fuel.
In what follows, we investigate two different mechanisms for removing
the cold interstellar medium of infalling satellites.

\subsubsection{Ram-Pressure Stripping}
Following \citet{gunn72}, we estimate the ram pressure ($P_{\rm ram}$)
as:
\begin{equation}
P_{\rm ram} \sim \rho_{\rm halo} V_{\rm sat}^{2} \; ,
\label{eq:ram}
\end{equation}
where $\rho_{\rm halo}$ is the density of the host's gas halo and
$V_{\rm sat}$ is the velocity of the satellite galaxy with respect to
the host's frame of reference, or more precisely the local reference
frame of the host's gas halo in the immediate vicinity of the
infalling satellite galaxy.
As shown in Equation~\ref{eq:ram}, the ram pressure experienced by an
infalling satellite is dependent on the local environment --- thus,
the properties of the host system, in particular its dark matter halo
mass, which plays a critical role in setting $\rho_{\rm halo}$ and
$V_{\rm sat}$.
As described in Section~\ref{subsec:rhoV}, we utilize $N$-body
simulations and observations of the Local Group and similar systems to
inform our selection of these global environmental parameters,
applying average values of $\rho_{\rm halo}$ and $V_{\rm sat}$ to all
infalling satellites in our analysis.
The adopted values for these parameters, along with uncertainties or
biases that their selection introduces, are discussed directly in
Section~\ref{subsec:drhoV}.

Assuming a spherical mass profile for an infalling satellite, the
gravitational restoring force per unit area is given by:
\begin{equation}
P_{\rm restore} \sim \Sigma_{\rm gas}(r) \frac{G \, M(r)}{r^2}
\; ,
\label{eq:restore}
\end{equation}
where $\Sigma_{\rm gas}(r)$ is the surface density of the cold gas to
be stripped from the satellite and $M(r)$ is the total satellite mass
interior to the radius $r$.
As shown in Equation~\ref{eq:restore}, the restoring pressure depends
exclusively on the properties of the infalling galaxy, varying from
one satellite system to the next, as it is accreted onto the parent
halo.
To model the properties of a representative sample of infalling
satellites, we utilize observational data for a broad collection of
nearby galaxies, including mass modeling to infer the local
gravitational potential on a system-by-system basis (see
Section~\ref{subsec:sigM}).

For an infalling satellite, the degree to which ram pressure is able
to strip its ISM is determined by the relative magnitude of the two
pressures ($P_{\rm ram}$ versus $P_{\rm restore}$), such that
stripping will occur beyond a radius $r$ (within the satellite) if
\begin{equation}
\rho_{\rm halo} V_{\rm sat}^{2} \ \gt \ \Sigma_{\rm gas}(r)
\frac{G \, M(r)}{r^2}  \; .
\label{eq:test}
\end{equation}
Throughout this work, we define $R_{\rm strip}$ as the innermost
radial distance at which this inequality holds. 
Inside $R_{\rm strip}$, the restoring pressure is able to resist
stripping, while ram pressure dominates beyond this radius.

\subsubsection{Turbulent Viscous Stripping}

The interaction at the interface of the ISM and the CGM can result in
the growth of Kelvin-Helmholtz (K-H) instabilities due to the relative
motion between the two phases. 
This will allow turbulent viscous stripping to remove the outer
regions of the ISM when the gravitational restoring force is
sufficiently small. Perturbations with wavenumber, $k$, are unstable
if they meet the following criteria \citep{murray93, mori00}:
\begin{equation}
k \gt g \, \frac{\rho_{\rm gas}^{2} - \rho_{\rm halo}^{2}}{\rho_{\rm gas}
  \, \rho_{\rm halo} \, V_{\rm sat}^{2}} \; ,
\label{eq:k}
\end{equation}
where $g$ is the gravitational restoring force at the ISM-CGM
interface.

Previous studies of turbulent viscous stripping find that the dominant
wavelength is set by the size of the cold gas region ($R_{\rm ISM}$),
such that $k = 2\pi / R_{\rm ISM}$ \citep{nulsen82, murray93}.  In our
analysis, we make the assumption that $\rho_{\rm halo} \ll \rho_{\rm
  gas}$, which is undoubtedly true for gas-rich dwarfs accreted into
the Local Group (or similar environments). Plugging these
approximations into Equation~\ref{eq:k}, leads to the following
inequality of the same form as Equation~\ref{eq:test}:
\begin{equation}
  \rho_{\rm halo} V_{\rm sat}^{2} \ \gt \frac{G \, M_{0} \, \bar{\rho}_{\rm
      gas}}{2 \pi \, R_{\rm ISM}}  \; ,
\label{eq:kh_check}
\end{equation}
where $\bar{\rho}_{\rm gas}$ is the average ISM density inside $R_{\rm
  ISM}$, and $M_{0}$ is the total restoring mass inside $R_{\rm
  ISM}$. If this inequality is true, then turbulent viscous stripping
will proceed and the outer layers of the ISM will be removed. When
this inequality is false, the gravitational restoring pressure is able
to stabilize the outer layers of the ISM against the K-H instabilities.

When turbulent viscous stripping is able to proceed, the rate at which
the ISM is removed will determine how much gas is stripped and
ultimately whether the reservoir for star formation will be
significantly depleted. The rate of total gas mass loss ($\dot{M}$) is
given in slightly different forms throughout the literature
\citep[e.g.][]{nulsen82, mori00, roediger05}. In this work, we adopt
the following approximation from \citet{roediger05}:
\begin{equation}
  \dot{M} \approx 20 \, \left (\frac{R_{\rm ISM}}{20 \, {\rm kpc}} \right )^{2}
  \left (\frac{n_{\rm halo}}{10^{-3} \, {\rm cm}^{-3}} \right ) \left
    (\frac{V_{\rm sat}}{1000 \, {\rm km} \, {\rm s}^{-1}} \right )
\frac{\msun}{\rm yr} \; .
\label{eq:mloss}
\end{equation}
The details regarding how Equations~\ref{eq:kh_check} and
\ref{eq:mloss} are used to determine the fraction of ISM removed from
an infalling dwarf galaxy are discussed further in
Section~\ref{subsec:frac}.

\begin{figure}
 \centering
 \hspace*{-0.07in}
 \includegraphics[width=3.35in]{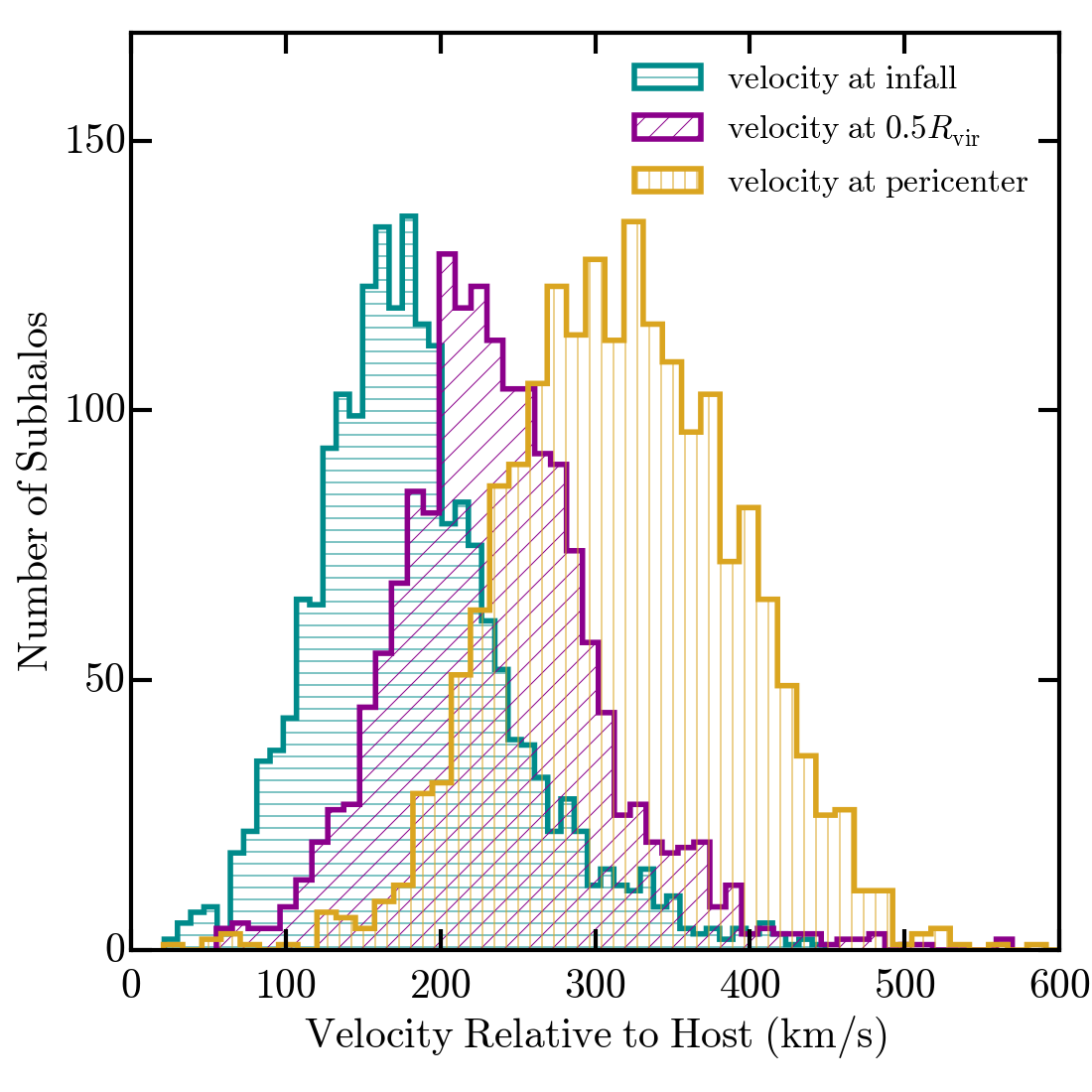}
 \caption{The distribution of subhalo velocities (relative to their
   parent halo) in the ELVIS suite of Local Group simulations for all
   subhalos that reside inside the virial radius at $z = 0$. The cyan,
   magenta, and gold histograms correspond to
   subhalo velocities measured when crossing the virial radius
   (i.e.~infall), at $0.5~R_{\rm vir}$, and at pericenter,
   respectively. For Milky Way-like systems, the typical satellite
   velocity (relative to the host's halo gas) is roughly
   $200-400$~km~s$^{-1}$ at the time of quenching. In our analysis, we
   adopt a fiducial value of $V_{\rm sat} = 300$~km~s$^{-1}$.}
 \label{fig:vel}
\end{figure}

\subsection{Estimating \boldmath$\rho_{\rm halo}$ and \boldmath$V_{\rm sat}$}
\label{subsec:rhoV}

The strength of the stripping force acting upon an infalling satellite is
primarily set by the density of the host's halo gas along with the
relative velocity of the satellite (see Eq.~\ref{eq:ram} and Eq.~\ref{eq:kh_check}).
Within the Milky Way, a variety of indirect probes point towards halo
gas densities of $\sim 10^{-4}$~cm$^{-3}$ for the hot
($T\sim10^{6}$~K) component \citep[e.g.][]{weiner96, snez02, fox05,
  grcevich09, salem15}.
Moreover, both observed X-ray emission and pulsar dispersion
measurements in the Milky Way are consistent with a cored hot halo
distribution with a density of $> 10^{-4}$~cm$^{-3}$
extending to radial distances of $\sim100$~kpc \citep[][see also
\citealt{anderson10, gupta12, miller13, miller15, fsm16}]{fang13}.

While the Milky Way's hot halo component is clearly important with
regard to stripping in the Local Group, it is the density
of the host's halo gas -- across all temperatures -- that dictates the
strength of the stripping force.
When folding in cooler phases of the circumgalactic medium, recent
studies of nearby massive galaxies, comparable to the Milky Way, find
halo gas densities of $\sim10^{-3.5}$~cm$^{-3}$ extending to
at least $\gtrsim0.25~R_{\rm vir}$ \citep{tumlinson13, werk14, fsm16}.
These results are also supported by the latest analysis of the CGM
surrounding M31 using quasar absorption-line spectroscopy, which finds
evidence for a massive and extended gas halo \citep{lehner15}.
Related studies targeting more massive, high-$z$ systems find
extended, high-density reservoirs of cool ($10^{4}$~K) halo gas
reaching out to large fractions of the virial radius \citep{lau15}.
In accordance with these recent results, we assume a fiducial value
for the host halo density ($n_{\rm halo}$) of $10^{-3.5}$~cm$^{-3}$,
where $\rho_{\rm halo} =\mu n_{\rm halo} m_{\rm H\scalebox{0.45}{\rm
    I}}$. For the purposes of this analysis, the mean molecular
weight, $\mu $, is set to $1$.

Given the significant uncertainties in the observed halo densities of
the Milky Way and M31, we explore how both our calculations of
instantaneous ram-pressure stripping and continuous viscous stripping
depend on this adopted value of
$n_{\rm halo}$ in Section~\ref{subsec:drhoV}.
Throughout our analysis, we make no assumptions regarding the radial
profile of the halo gas. 
Given the expected orbits of infalling satellite populations and the
existing constraints on the quenching timescale measured relative to
infall, however, there are relevant constraints regarding the extent
of the CGM, which we discuss in Section~\ref{subsec:BigPic}.

To estimate the relative velocity of a satellite system in relation to
the host's halo gas ($V_{\rm sat}$), we study the distribution of
subhalo velocities within the Exploring the Local Volume In
Simulations (ELVIS) suite of $48$ high-resolution, dissipationless
simulations of Milky Way-like halos \citep{gk14}.
The ELVIS suite includes $24$ isolated halos as well as $12$
mass-matched Local Group-like pairs, simulated within high-resolution
uncontaminated volumes spanning $2-5$~Mpc in size using a particle
mass of $1.9 \times 10^{5}~\msun$ and a Plummer-equivalent force
softening of $\epsilon = 141$~physical parsecs. Within the
high-resolution volumes, the halo catalogs are complete down to
$M_{\rm halo} > 2 \times 10^{7}~\msun$, $V_{\rm max} > 8$~km~s$^{-1}$,
$M_{\rm peak} > 6 \times 10^{7}~\msun$, and $V_{\rm peak} > 12$~km
s$^{-1}$ -- thus more than sufficient to track the evolution of halos
hosting the Local Group dwarf population.

From ELVIS, we select subhalos corresponding to satellites with
stellar masses of $10^{6}-10^{9}~\msun$ --- i.e.~halo masses of
$10^{9.7}-10^{11.2}~\msun$ following the stellar mass-halo mass (SMHM)
relation of \citet{gk14}. 
We sample the velocities of these subhalos at the time of infall
(i.e.~crossing $R_{\rm vir}$), at $0.5~R_{\rm vir}$, and at
pericenter. In Figure~\ref{fig:vel}, we show the distribution of
subhalo velocities (relative to that of their host dark matter halo)
at each of these distances. 
As expected, the average velocity of the subhalo population increases
from infall towards pericenter, with a mean velocity of $183$, $237$,
and $318$~km~s$^{-1}$ at $R_{\rm vir}$, $0.5~R_{\rm vir}$, and $R_{\rm
  peri}$, respectively.\footnote{Both the subhalo velocity and
  pericenter distributions show no dependence on subhalo mass,
  consistent with the idea that the host potential is the primary
  driver of these subhalo properties.}
To increase the precision at which we are able to measure the position
of pericentric passage and the velocity at pericenter for each subhalo,
we 3-d spline interpolate the position and velocity information for
all subhalos across the $75$ simulation snapshots in ELVIS.
While this interpolation scheme achieves a time resolution of
$\sim20$~Myr, our resulting measurements of $R_{\rm peri}$ and $V_{\rm
  peri}$ are likely somewhat over- and under-estimated, respectively.

Typically, studies of instantaneous ram-pressure stripping assume the
satellite velocity at pericenter, where stripping is expected to be
greatest.
Quenching (i.e.~stripping) when a satellite reaches a radial distance
of $\sim0.5~R_{\rm vir}$ while on first infall, however, is consistent
with the inferred quenching timescales for the Local Group satellite
population \citep{fham15}. 
Herein, we compromise between these two scenarios, adopting a fiducial
satellite velocity ($V_{\rm sat}$) of $300$~km~s$^{-1}$. In
Section~\ref{subsec:drhoV}, we explore how our results depend upon
this choice of $V_{\rm sat}$.

\begin{figure}
 \centering
 \hspace*{-0.15in}
 \includegraphics[width=3.4in]{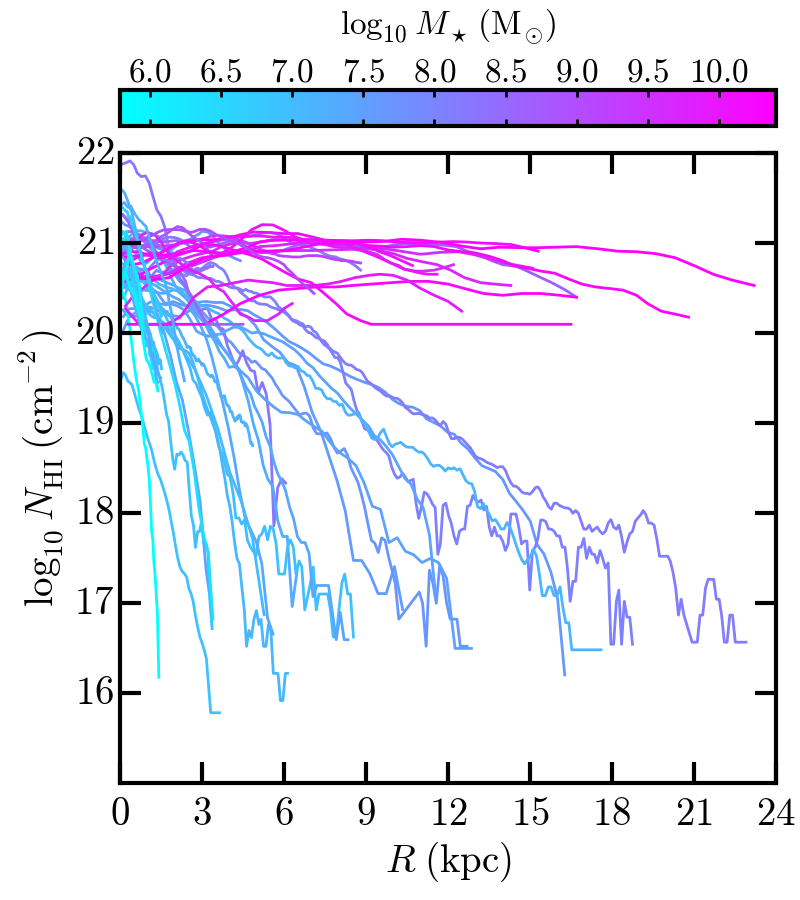}
 \caption{The observed H{\scriptsize I} surface density profiles for
   our sample of $66$ nearby field galaxies from THINGS, Little
   THINGS, and SHIELD, color-coded according to the stellar mass of
   each system. While there is substantial scatter in profile shape
   from object to object, the lower-mass dwarfs are preferentially
   less extended than their more massive counterparts.}
 \label{fig:h1}
\end{figure}

\subsection{Estimating \boldmath$\Sigma_{\rm gas}(r)$ and \boldmath$M(r)$}
\label{subsec:sigM}

As shown in Equation~\ref{eq:restore}, the gravitational restoring
pressure is dependent upon the properties of the infalling satellite,
specifically the gas surface density and total mass profiles.
To estimate these parameters for a representative sample of infalling
satellites, we utilize existing observations of $66$ nearby,
star-forming dwarf galaxies from the THINGS, Little THINGS, and SHIELD
data sets (\citealt{walter08, hunter12, cannon11}; McNichols et al.~in
prep; Teich et al.~in prep). 
This sample is dominated by isolated (or ``field'') systems, for which
the ISM is largely unaltered by previous interactions with a more
massive host system --- i.e.~ideal candidates to test the
effectiveness of ram-pressure stripping as a quenching mechanism.
While the satellites of the Milky Way and M31 were primarily accreted
at $z \sim 0.5-1$ \citep{wetzel15a, fham15}, our sample of nearby
galaxies is expected to be similar in cold gas (specifically
H{\scriptsize I}) content to similar systems at intermediate redshift
\citep{popping15, somerville15}.

For galaxies drawn from Little THINGS and SHIELD, we infer the stellar
mass of the system using the published $V$-band absolute magnitudes
\citep{hunter12, hauerberg15} and assuming a mass-to-light ratio of
unity, which is roughly consistent with the expectations for a
$\sim1$~Gyr-old simple stellar population following a Salpeter initial
mass function \citep[e.g.][]{maraston98}.
For those systems selected from THINGS, we utilize the stellar mass
estimates of \citet{leroy08}, which are derived from {\it Spitzer}
3.6$\mu$m imaging assuming a $K$-band mass-to-light ratio of $0.5$.
As discussed further in Section~\ref{subsec:dMr}, uncertainties in the
measured stellar masses for our sample have little impact on the
quantitative or qualitative results of our analysis.
For the $12$ galaxies in the SHIELD sample, the resulting stellar mass
estimates are in relatively good agreement with those derived from
stellar population fits to multi-band {\it Hubble Space Telescope}
photometry, with a typical offset (to higher masses) of $0.37$ dex
\citep{mcquinn15}.
Altogether, the sample of $66$ field systems spans a broad range in
stellar mass, from $\sim10^{6}-10^{11}~\msun$, covering the mass
regime where \citet{fham15} find evidence for a change in the dominant
satellite quenching mechanism and where stripping is presumed to
become effective.

For each of the galaxies in our sample, we utilize the published
H{\scriptsize I} surface density profiles from the THINGS, Little
THINGS, and SHIELD projects, scaled by a factor of $1.36$ to account
for helium (\citealt{leroy08, hunter12, teich15}; Teich et al.~in
prep).
For a typical low-mass galaxy, the cold gas component is largely
dominated by atomic (versus molecular) gas \citep[e.g.][]{popping14,
  boselli14}, such that $\Sigma_{\rm H\scalebox{0.45}{\rm I}}$
provides a robust estimate of the ISM surface density and thus the
efficacy of both ram-pressure and turbulent viscous stripping.
As shown in Figure~\ref{fig:h1}, the $\Sigma_{\rm H\scalebox{0.45}{\rm
    I}}$ profiles for our sample exhibit significant variation in
shape, with more massive systems having preferentially more extended
H{\scriptsize I} surface density profiles.
While the depth of the H{\scriptsize I} observations varies from
object to object in our sample, the THINGS, Little THINGS, and SHIELD
measurements are sensitive to the bulk of the atomic gas component,
such that any undetected low-density gas at large radii would have a
negligible impact on our stripping calculations. 

To determine the mass profile, $M(r)$, for each galaxy in our sample,
we infer the total dark matter halo mass according to the stellar
mass-halo mass relation of \citet{gk14} and assume an NFW density
profile \citep{nfw97} with a concentration given by the $c-M$ relation
of \citet{klypin11}.
While this methodology neglects contributions to the mass profile from
the baryonic component, these are relatively modest at these mass
scales (i.e.~$\mstar < 10^{11}~\msun$), as illustrated in
Section~\ref{subsec:dMr}.
Recognizing current uncertainties in the dark matter density profiles
of low-mass galaxies \citep[e.g.][]{moore94, deblok01, bk11, bk12}, we
also employ mass profiles derived from dynamical modeling of the
observed H{\scriptsize I} kinematics for a subset of our systems,
including NFW fits to the THINGS and Little THINGS samples from
\citet{deblok08} and \citet{oh15} as well as fits to a Burkert profile
\citep{burkert95} from \citet{pace16}.


\begin{figure*}
 \centering
 \hspace*{-0.2in}
 \includegraphics[width=5.0in]{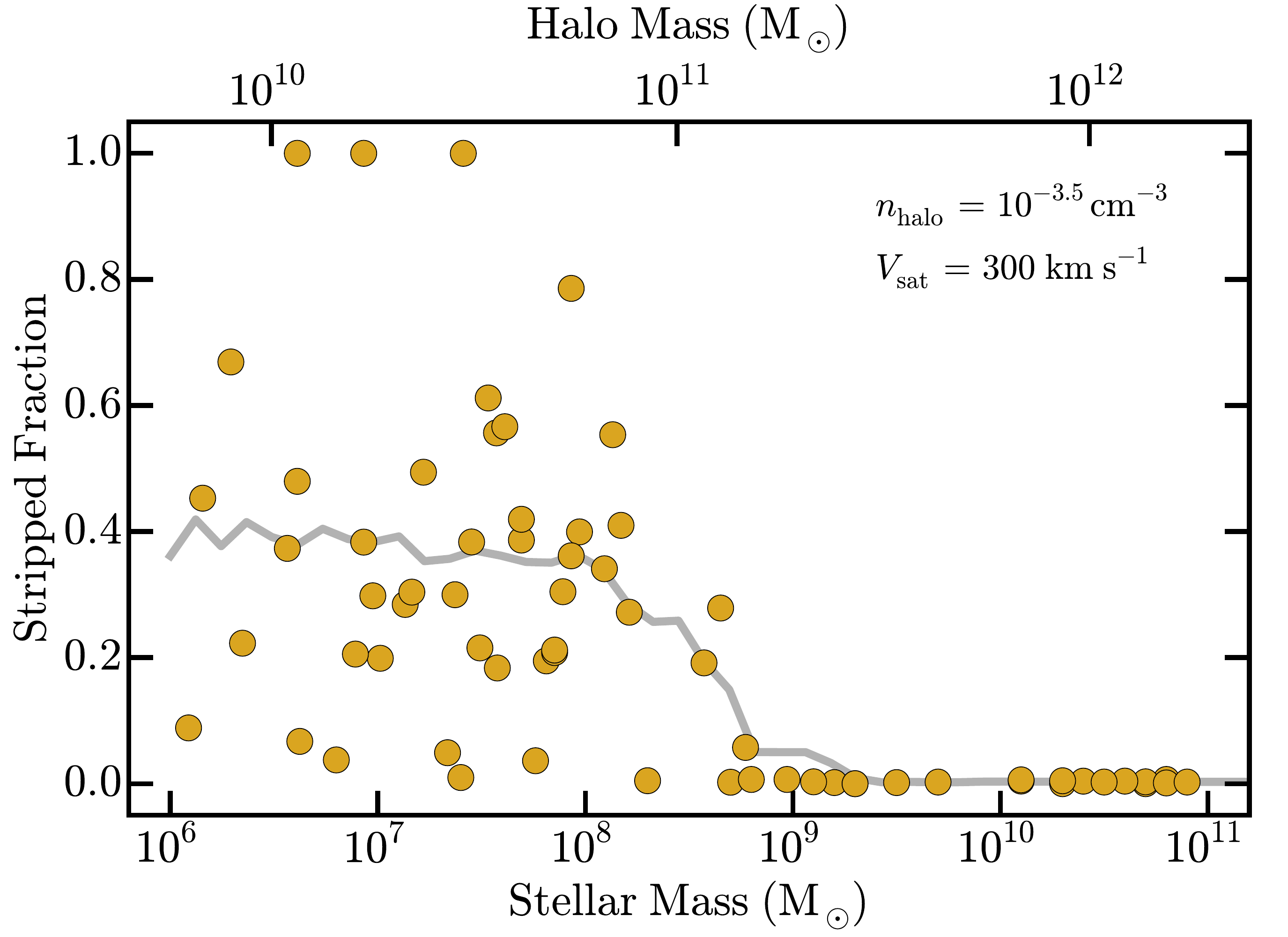}
 \caption{The fraction of H{\scriptsize I} gas stripped ($f_{\rm
     stripped}$) via ram pressure as a function of stellar mass for
   our sample of $66$ dwarf galaxies, assuming a host halo gas density
   of $n_{\rm halo} = 10^{-3.5}~{\rm cm}^{-3}$ and a satellite
   velocity of $V_{\rm sat}=300~{\rm km}~{\rm s}^{-1}$. The grey solid
   line corresponds to the mean $f_{\rm stripped}$ computed in a
   sliding bin of width $0.6$ dex in stellar mass. At stellar masses
   greater than roughly $10^{9}~\msun$, we find that satellite systems
   are unaffected by ram pressure in a Milky Way-like environment. At
   $\mstar \lesssim 10^{8.5}~\msun$, however, ram pressure is
   increasingly effective, with infalling systems typically having
   $\sim 40\%$ of their cold gas stripped. It is worth noting that the
   scatter in the stripped fraction at fixed stellar mass is driven
   entirely by the variation in the H{\scriptsize I} surface density
   profiles. For reference, we include
   the corresponding halo mass for each system, as inferred via the
   stellar mass-halo mass relation of \citet{gk14}.}
\label{fig:fid}
\end{figure*}


\subsection{Measuring the Stripped Fraction (\boldmath$f_{\rm stripped}$)}
\label{subsec:frac} 
Using the H{\scriptsize I} surface density profiles presented in
Figure~\ref{fig:h1}, we determine the fraction of H{\scriptsize I} gas
stripped from each satellite in our sample, given an assumed host halo
density ($\rho_{\rm halo}$), satellite velocity ($V_{\rm sat}$), and
satellite mass profile ($M(r)$).
Throughout our analysis, we first determine the amount of ISM removed
via instantaneous ram-pressure stripping, then we allow turbulent
viscous stripping to proceed for up to $1$~Gyr.

\subsubsection{Ram-Pressure Stripping}
First, the satellite experiences ram pressure stripping, which is
generally assumed to coincide with either initial infall or pericentric
passage. As discussed in Section~\ref{subsec:af},
Equation~\ref{eq:test} specifies the radial distance, measured from
the center of each satellite, at which ram pressure exceeds the
gravitational restoring pressure ($R_{\rm strip}$).
By integrating the H{\scriptsize I} surface density profile beyond
this radius, we compute the fraction of gas stripped from each
satellite as
\begin{equation}
f_{\rm stripped} = \frac{\int_{R_{\rm strip}}^{R_{\rm max}} \Sigma_{\rm gas}(r)\,
  r\, {\rm d}r}{\int_{0}^{R_{\rm max}} \Sigma_{\rm gas}(r)\, r\, {\rm d}r}  \; ,
\label{eq:frac}
\end{equation}
where $R_{\rm max}$ is the outermost radial distance at which
H{\scriptsize I} is detected.
Here, the numerator corresponds to the gas mass that is stripped from
the satellite after it interacts with the CGM of the host. The
denominator is the total gas mass that resides in the system in the
absence of any environmental effects (i.e.~prior to infall).
The stripped fraction in this scenario is the amount of gas removed in
a single, instantaneous interaction between the infalling satellite
and the host halo.

As discussed in Section~\ref{subsec:sigM}, the adopted definition for
$R_{\rm max}$ leads to an underestimate of the stripped fraction, as
low-density gas at large galactocentric radii is unaccounted for in
our analysis. However, given the sensitivities of the THINGS, Little
THINGS, and SHIELD H{\scriptsize I} maps, any H{\scriptsize I}
component at large radii contributes minimally to the total atomic gas
mass, such that the resulting impact on $f_{\rm stripped}$ should be
negligible.

\subsubsection{Turbulent Viscous Stripping}
After estimating $f_{\rm stripped}$ as a result of ram-pressure
stripping, we compute the corresponding fraction of gas removed due to
turbulent viscous stripping over a maximum timespan of $1$~Gyr.
First, we test whether the ISM and CGM interface conditions located at
$R_{\rm strip} $ are susceptible to viscous stripping via
Equation~\ref{eq:kh_check}.
If true, we determine the gas mass lost during a $100$~Myr
interval, $M_{\rm viscous}$, using Equation~\ref{eq:mloss}.
$M_{\rm viscous}$ is then uniformly removed from the outermost regions of
the H{\scriptsize I} surface density profile, leading to a new $R_{\rm
  strip} $ in addition to a new value of both $\bar{\rho}_{\rm
  gas}$ and $M_{0}$.
The ISM-CGM conditions are then reevaluated allowing the gas removal
process to continue if Equation~\ref{eq:kh_check} is still true. 
We repeat this process for up to $1$~Gyr, leading to a total gas mass
lost via turbulent viscous stripping. 
This additional gas mass is added to the gas which was initially
removed via ram pressure to get a total gas mass lost as a result of
stripping.

We limit the timespan on which turbulent viscous stripping occurs, so
as to roughly match the measured quenching timescale for Local Group
satellites \citep[i.e.~$\lesssim 2$~Gyr,][]{fham15}.
Approximately $50\%$ (or $55\%$) of the subhalo population in ELVIS
reaches pericenter (or $0.5~R_{\rm vir}$) within $\sim1.5$~Gyr (or
$\sim1$~Gyr) of infall, where instantaneous ram-pressure stripping is
assumed to occur.
A further $1$~Gyr of turbulent viscous stripping therefore yields a
typical quenching time (relative to infall) in rough agreement with
the expectations of \citet{fham15} and \citet{wetzel15b}.

\section{Results}
\label{sec:results}

In Figure~\ref{fig:fid}, we show the fraction of H{\scriptsize I} gas
ram-pressure stripped from our sample of star-forming galaxies (with
no viscous stripping), assuming a host halo density of $n_{\rm halo} =
10^{-3.5}$~cm$^{-3}$ and a satellite velocity of $V_{\rm sat} =
300$~km~s$^{-1}$.
For satellite systems with stellar masses greater than
$\sim~10^{9}~\msun$, ram pressure is unable to strip the interstellar
medium, consistent with the long quenching timescales inferred by
\citet{wheeler14} and \citet{fham15} at this mass regime.
At $\mstar \sim 10^{8-9}~\msun$, however, we find that ram pressure
begins to overcome the local gravitational restoring force, such that
a significant fraction of the cold gas reservoir is removed from a
typical infalling satellite at $\mstar < 10^{8}~\msun$ ($<\!f_{\rm
  stripped}\!> \sim 40\%$).
While there is considerable scatter in the efficacy of ram-pressure
stripping at low masses, our fiducial model for a Milky Way-like
system qualitatively reproduces the critical mass scale for quenching
at $\sim10^{8}~\msun$, such that ram-pressure stripping is a viable
candidate to be the dominant quenching mechanism at low satellite
stellar masses.
In Section~\ref{subsec:dMr} and \ref{subsec:drhoV}, we explore how
this result depends on the specific parameters adopted in our fiducial
model (i.e.~$M(r)$, $n_{\rm halo}$, and $V_{\rm sat}$). Additionally,
in Section~\ref{subsec:viscous}, we discuss how the inclusion of
turbulent viscous stripping impacts these results.


\begin{figure*}
 \centering
 \hspace*{-0.25in}
   \includegraphics[width=7.7in]{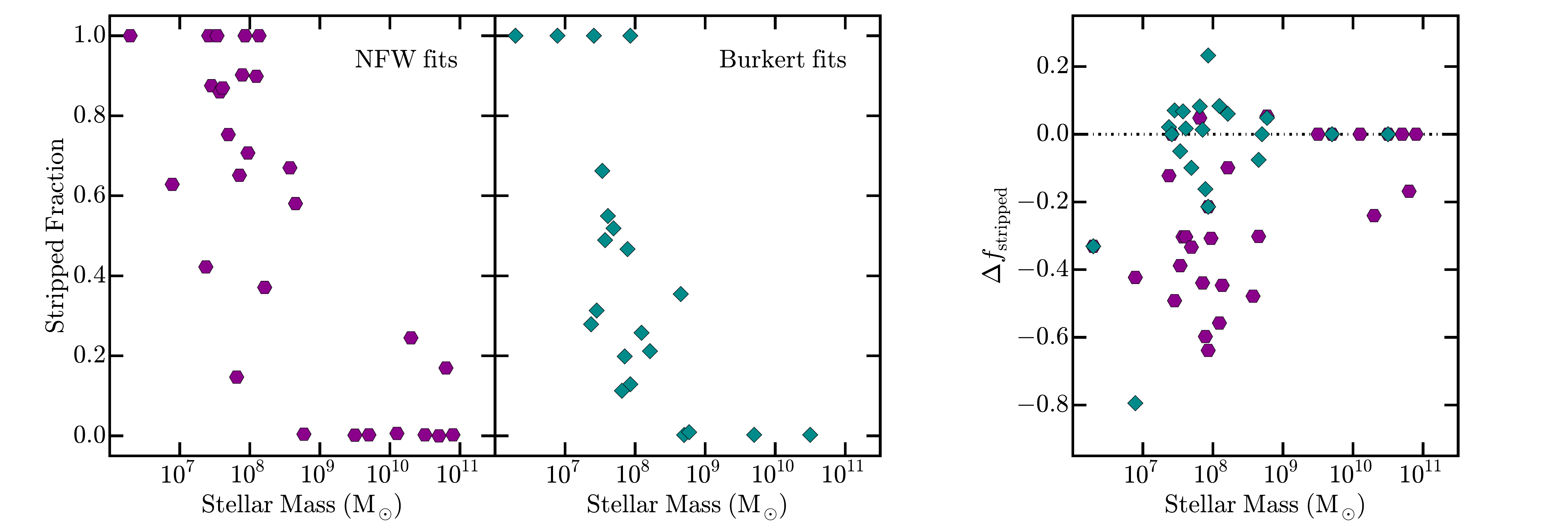}
   \caption{(\emph{Left}): the fraction of H{\scriptsize I} gas
     ram-pressure stripped ($f_{\rm stripped}$) as a function of
     stellar mass for the subset of systems with mass profiles
     determined via dynamical fits to an NFW profile \citep{deblok08,
       oh15} and to a Burkert profile \citep{pace16}. As in
     Fig.~\ref{fig:fid}, we assume a halo gas density of $n_{\rm
       halo}=10^{-3.5}~{\rm cm}^{-3}$ and a satellite velocity of
     $V_{\rm sat}=300~{\rm km}~{\rm s}^{-1}$. (\emph{Right}): the
     difference in the fraction of H{\scriptsize I} gas stripped
     relative to the corresponding result (see Fig.~\ref{fig:fid})
     assuming our fiducial mass profile inferred via the stellar
     mass-halo mass relation of \citet{gk14}: ${\Delta}f_{\rm
       stripped} = f_{\rm stripped, fid} - f_{\rm stripped, dyn}$. In
     general, the mass profiles inferred from dynamical modeling favor
     cored halos, such that stripping is more efficient relative to
     our fiducial model. Moreover, while there is significant scatter
     from galaxy to galaxy based upon the assumed mass profile, the
     qualitative results are universal with ram-pressure stripping
     becoming increasingly effective at $\mstar<10^{8-9}~\msun$. }
 \label{fig:mass}
\end{figure*}

\subsection{Dependence on \boldmath$M(r)$}
\label{subsec:dMr}

The assumed dark matter density profile for each galaxy in our sample
is critical in determining the strength of the local gravitational
restoring pressure (see Eq.~\ref{eq:restore}) and thus the degree to
which ram pressure is able to strip the infalling satellite's ISM.
In our fiducial model, the mass profiles, $M(r)$, are determined using
the stellar mass-halo mass (SMHM) relation of \citet{gk14}, assuming
an NFW density profile.
At low masses, however, less-cuspy dark matter profiles are typically
favored and there are large uncertainties in the slope of the stellar
mass-halo mass relation (and its scatter, \citealt{gk16}).

To explore how our estimates of $f_{\rm stripped}$ depend on the
assumed mass profile, $M(r)$, we utilize alternative mass profiles
derived from dynamical fits to the observed H{\scriptsize I}
kinematics for a subset of the systems in our sample. In particular,
we utilize the NFW fits to the THINGS and Little THINGS velocity
fields from \citet{deblok08} and \citet{oh15}, respectively. 
For $21$ galaxies, we also employ Burkert profile fits to the
H{\scriptsize I} kinematics from \citet{pace16}.

The two left most panels of Figure~\ref{fig:mass} show the fraction of
atomic gas stripped for the subset of objects with dynamical mass
estimates, assuming a host halo density of $n_{\rm halo} =
10^{-3.5}$~cm$^{-3}$ and a satellite velocity of $V_{\rm
  sat} = 300$~km~s$^{-1}$. 
Qualitatively, the dependence of $f_{\rm stripped}$ on satellite
stellar mass is very similar to that shown in Fig.~\ref{fig:fid} for
our fiducial model, which employs mass profiles inferred from the SMHM
relation of \citet{gk14}.
At $\mstar < 10^{8-9}~\msun$, ram-pressure stripping becomes
increasingly effective. 
However, the stripped fractions calculated using the dynamical mass
profile fits are, on average, slightly greater relative to those
produced by our fiducial model. 
This effect is evident in the far right panel of
Figure~\ref{fig:mass}, which shows the difference in the stripped
fraction for each object as we vary the restoring mass profile.
For the dynamical fits to an NFW and Burkert profile (magenta hexagons
versus cyan diamonds), we find a mean difference in $f_{\rm stripped}$
of $-0.32$ and $-0.15$, respectively. 
On average, the dynamical fits lead to greater stripped fractions,
consistent with these objects being hosted by less-concentrated (or
lower-mass) dark matter halos at fixed stellar mass. While there is
not perfect agreement between the different halo mass estimators, by
adopting the SMHM relation as our fiducial method, we are likely
underestimating the stripped fraction and thus providing a
conservative estimate of the effectiveness of ram-pressure stripping.


\begin{figure*}
 \centering
 \hspace*{-0.5in}
  \includegraphics[width=7.6in]{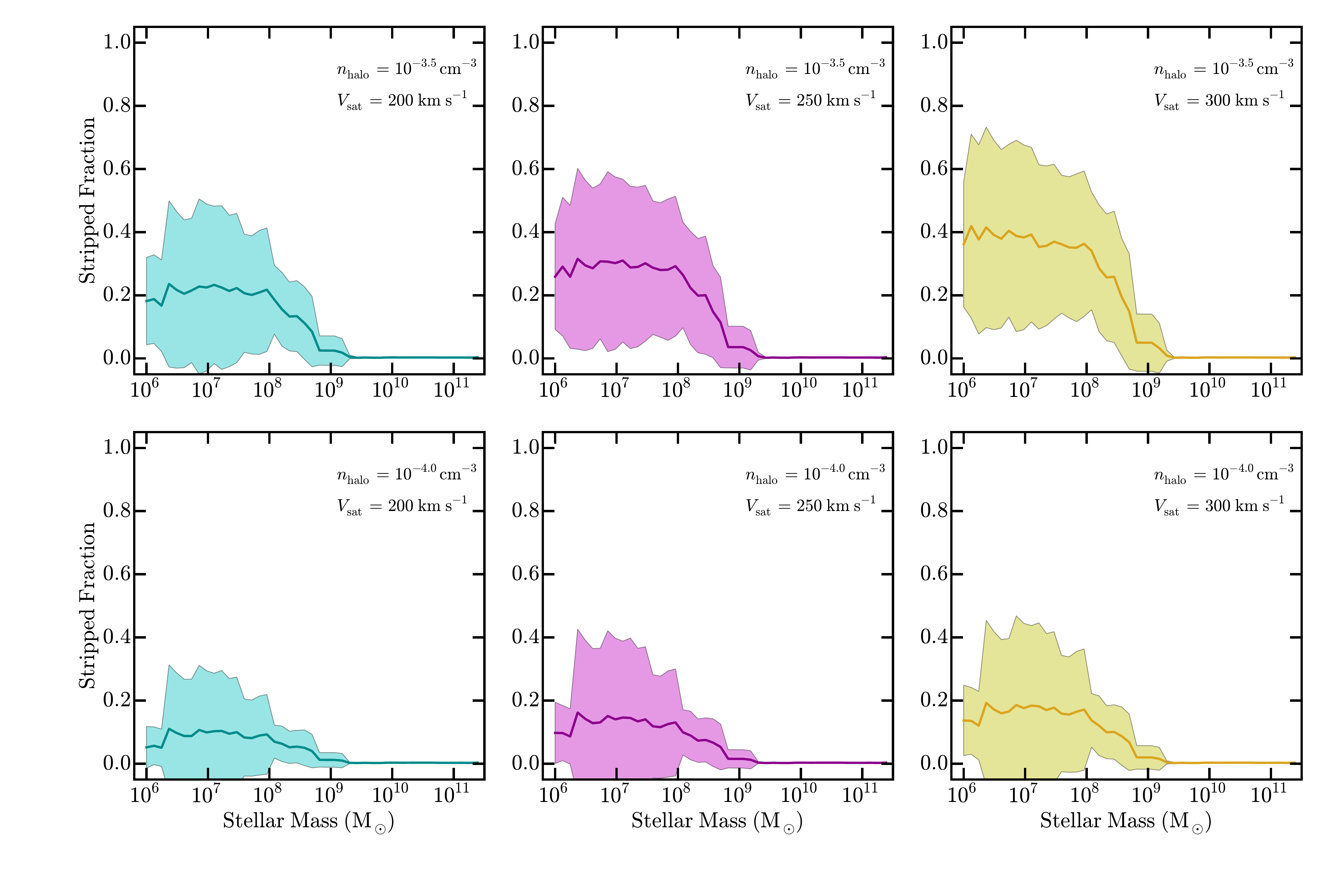}
  \caption{The fraction of H{\scriptsize I} gas stripped ($f_{\rm
      stripped}$) via ram-pressure 
    stripping as a function of satellite stellar mass for our sample
    of $66$ dwarf galaxies. The solid line in each panel gives the
    mean $f_{\rm stripped}$ in a sliding bin of width $0.6$ dex in
    stellar mass, with the shaded region tracing the corresponding
    $1\sigma$ scatter. In the \emph{top} and \emph{bottom} rows, we
    assume a host halo gas density of $n_{\rm
      halo}=10^{-3.5}$~cm$^{-3}$ and $10^{-4.0}$~cm$^{-3}$,
    respectively. From \emph{left} to \emph{right}, the satellite
    velocity varies from $200$ (cyan) to $250$ (magenta) to
    $300$~km~s$^{-1}$ (gold). While the efficiency of ram-pressure
    stripping depends on the assumed properties of the host halo, such
    that $<\!f_{\rm stripped}\!>$ ranges from $\sim10-40\%$, the
    satellite stellar mass where ram-pressure stripping becomes
    significant is universally $<10^{9}~\msun$.}
 \label{fig:MW}
\end{figure*}


\subsection{Dependence on \boldmath$n_{\rm halo}$ and
  \boldmath$V_{\rm sat}$}
\label{subsec:drhoV}

In addition to the uncertainty in the restoring mass profile, the
amount of cold gas stripped from each dwarf is highly dependent on the
properties of the host system (i.e.~$n_{\rm halo}$ and $V_{\rm sat}$).
The density of the host gas halo, for the Local Group in particular,
is relatively poorly constrained. 
To explore how variation in these global parameters will impact our
results, we measure the stripped fraction for our sample while varying
both the satellite velocity, $V_{\rm sat} =\left \{200, 250, 300\right
\}~{\rm km~s}^{-1}$, and the density of the halo gas, $n_{\rm halo} =
\{10^{-4.0}, 10^{-3.5}\}~{\rm cm}^{-3}$.

Figure~\ref{fig:MW} shows the mean and $1\sigma$ scatter in the
stripped fraction as a function of satellite stellar mass for the
adopted variation in both the density of the CGM and the satellite
velocity relative to the frame of reference of the host.
Across the entire range of $V_{\rm sat}$ and $n_{\rm halo}$ explored,
ram-pressure stripping becomes effective at roughly the same critical
mass scale ($10^{8-9}~\msun$).
The efficacy of ram-pressure stripping at low masses, however, is
highly dependent on the chosen parameters for $n_{\rm halo}$ and
$V_{\rm sat}$.
For example, at $\mstar < 10^{9}~\msun$, the stripped fraction
decreases, on average, by $\sim0.15$ as the satellite velocity is
reduced from $300$ to $200$~km~s$^{-1}$ at fixed $n_{\rm halo}$.
Similarly, decreasing the host halo density from $n_{\rm halo} =
10^{-3.5}$~cm$^{-3}$ to $10^{-4}$~cm$^{-3}$ yields an average
reduction in $f_{\rm stripped}$ of $\sim0.2$ at fixed $V_{\rm sat}$.
The scatter in $f_{\rm stripped}$ associated with variation in $V_{\rm
  sat}$ is particularly noteworthy, given that in our analysis we
adopt a single value of $V_{\rm sat}$ for the entire satellite
population, thereby neglecting objects that have velocities greater
(or less) than this value. 
At pericentric passage, for example, roughly $60\%$ of subhalos in
ELVIS have velocities greater than $300~{\rm km~s}^{-1}$. This
fraction increases to $82\%$ and $96\%$ for $V_{\rm sat}$ values of
$250$ and $200~{\rm km~s}^{-1}$, respectively.


\begin{figure*}
 \centering
 \hspace*{-0.5in}
   \includegraphics[width=7.6in]{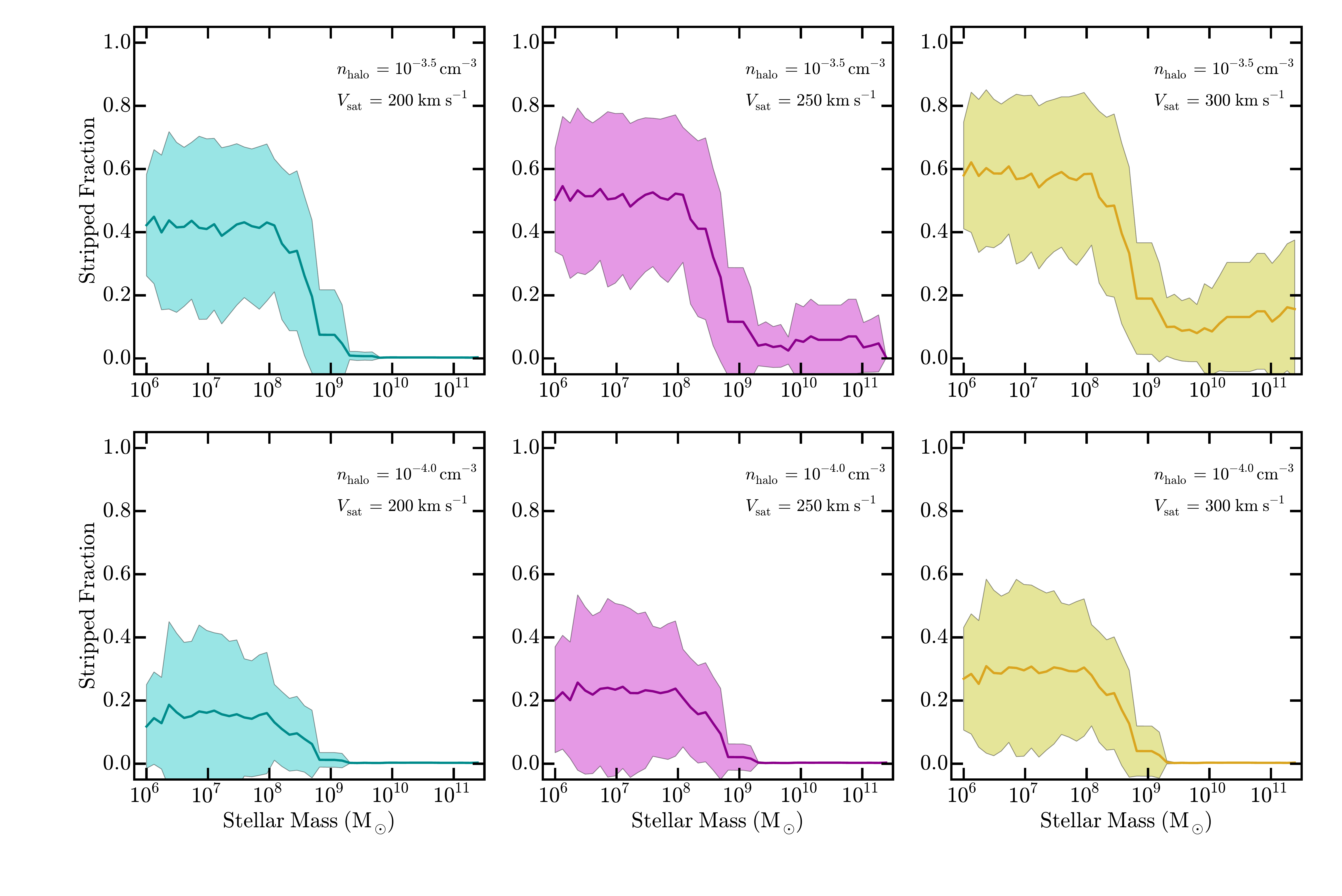}
   \caption{The fraction of H{\scriptsize I} gas stripped ($f_{\rm
       stripped}$) via ram-pressure \emph{and} turbulent viscous
     stripping as a function of satellite stellar mass for our sample
     of $66$ dwarf galaxies. The solid line in each panel gives the
     mean $f_{\rm stripped}$ in a sliding bin of width $0.6$ dex in
     stellar mass, with the shaded region tracing the corresponding
     $1\sigma$ scatter. In the \emph{top} and \emph{bottom} rows, we
     assume a host halo gas density of $n_{\rm
       halo}=10^{-3.5}$~cm$^{-3}$ and $10^{-4.0}$~cm$^{-3}$,
     respectively. From \emph{left} to \emph{right}, the satellite
     velocity varies from $200$ (cyan) to $250$ (magenta) to
     $300$~km~s$^{-1}$ (gold). Including both ram-pressure and
     turbulent viscous stripping, we find an increase in the fraction
     of stripped gas, such that the majority of gas is removed from
     low-mass satellites orbiting hosts with a halo gas density of
     $10^{-3.5}~{\rm cm}^{-3}$.} 
 \label{fig:MW_KH}
\end{figure*}


\subsection{Turbulent Viscous Stripping}
\label{subsec:viscous}

While ram-pressure stripping is effective at removing gas from
satellites below the critical quenching mass scale
($\lesssim~10^{8}~\msun$), it is only able to strip roughly half of
the cold gas reservoir on average.
As discussed in Section~\ref{sec:intro}, however, infalling satellite
systems are also subject to turbulent viscous stripping, which results
from K-H instabilities at the interface of the satellite's ISM and the
CGM.
To estimate the impact of this secondary stripping mechanism, we allow
turbulent viscous stripping to proceed for up to $1$~Gyr following the
initial ram-pressure stripping.

Figure~\ref{fig:MW_KH} shows the mean and $1\sigma$ scatter in the
stripped fraction due to both ram-pressure \emph{and} turbulent
viscous stripping as a function of satellite stellar mass, assuming
the same range of $V_{\rm sat}$ and $n_{\rm halo}$ values as in
Figure~\ref{fig:MW}.
In contrast to when ram-pressure acts alone, the inclusion of
turbulent viscous stripping at high satellite velocities ($V_{\rm sat}
\ge 250~{\rm km~s}^{-1}$) and host halo densities ($n_{\rm halo} \ge
10^{-3.5}$~cm$^{-3}$) yields non-zero stripped fractions for some
massive satellites.
As discussed in Section~\ref{sec:disc}, the amount of gas removed,
however, is relatively modest ($< \! f_{\rm stripped} \! > \; \lesssim
0.2$), in agreement with the observations of massive satellites in the
Local Group. 
At low stellar mass, the efficacy of stripping is notably increased
when including turbulent viscous effects, such that the typical
stripped fraction is roughly $1.5\times$ that produced by ram-pressure
stripping alone.
For a host halo density of $n_{\rm halo} = 10^{-3.5}$, the majority of
satellites in our sample are stripped of more than half of their cold
gas reservoirs.

\section{Discussion}
\label{sec:disc}

Recent studies of satellite galaxies in the local Universe find that
the efficiency of satellite (or environmental) quenching -- or the
timescale upon which it occurs following infall -- strongly depends on
the mass of the satellite system \citep{delucia12, wetzel13,
  wetzel15b, wheeler14, fham15}. For satellites with $\mstar \gtrsim
10^{8}~\msun$, the long quenching timescales inferred from the
relatively low observed satellite quenched fractions are consistent
with starvation as the dominant quenching mechanism \citep[][see also
\citealt{davies16}]{fham15}.
At stellar masses below a critical mass scale of $\mstar \lesssim
10^{8}~\msun$, however, the lack of quenched systems in the field
combined with the very high satellite quenched fractions observed in
the Local Group require a very short quenching timescale, consistent
with a physical process that acts on roughly the dynamical time, such
as ram-pressure stripping \citep{wetzel15b, fham15}. 
While stripping, and in particular ram-pressure stripping, is often
thought to be a possible factor in the dearth of star-forming dwarfs
in the Local Group \citep{einasto74, lin83, blitz00}, these recent
results provide possible benchmarks by which to measure stripping as
an active quenching mechanism.
In particular, does stripping become dominant at $\mstar \lesssim
10^{8}~\msun$, and is it strong enough to quench low-mass satellites
on a timescale of $\sim2$~Gyr in host systems such as the Milky Way or
M31?

\subsection{Reproducing the Critical Mass Scale for Satellite Quenching}
\label{subsec:critmass}

As first shown by \citet[][see also \citealt{slater14,
  phillips15a}]{wheeler14}, observations of galaxies in the Local
Volume point towards a remarkable shift in the efficiency of satellite
quenching below a satellite stellar mass of $\sim10^{8}~\msun$, such
that quenching at low masses proceeds relatively quickly following
infall.
Recent analysis of much larger samples of Milky Way-like hosts in deep
photometric datasets support this picture (Phillips et al.~in prep),
indicating a global critical mass scale for satellite quenching.
In agreement with this picture, we find that ram pressure begins to
overcome the local gravitational restoring force only in systems below
a stellar mass of $\sim10^{9}~\msun$.
Above this mass scale, dwarfs are largely resistant to ram-pressure
stripping, consistent with inefficient quenching via starvation.
At low masses, however, ram-pressure stripping is able to remove (at
least some of) the fuel for star formation from an infalling satellite
system, thus contributing to quenching and potentially driving the
change in quenching efficiency below $\sim10^{8}~\msun$.
Moreover, when including turbulent viscous stripping, this critical
scale for satellite quenching persists, with a dramatic increase in
stripping efficiency evident over the full range of $n_{\rm halo}$ and
$V_{\rm sat}$ values explored. 
Altogether, our results show that stripping naturally gives rise to a
critical scale for satellite quenching, consistent with observations
in the local Universe.

Unlike ram pressure, we find that at high halo densities and satellite
velocities turbulent viscous stripping is able to remove cold gas from
some infalling satellites in the high-mass regime
(i.e.~$>10^{9}~\msun$). 
The amount of gas removed, however, is relatively modest and in broad
agreement with observations of massive satellites in the Local Group.
For example, there is clear evidence for stripping of the LMC and SMC,
which are the only Milky Way satellites more massive than
$10^{8}~\msun$. However, the stripped gas that comprises the
Magellanic Stream and Leading Arm \citep{mathewson74}, may result from
tidal effects \citep[e.g.][]{lin77, besla10, besla12, guglielmo14}
versus ram-pressure or viscous stripping \citep[e.g.][]{md94,
  mastropietro05, salem15, hammer15}.
Given the uncertainties associated with the orbit of the LMC and SMC
and the physical origins of the stripped gas, it is difficult to make
a clear accounting of the stripped fraction for the two systems 
\citep[e.g.][]{donghia15}. However, the total H{\scriptsize I} gas
mass attributed to the Magellanic Stream and Leading Arm versus that
of the two Magellanic Clouds jointly is consistent with an average
stripped fraction for the two systems of $\lesssim25\%$, consistent
with our expectations at $\mstar>10^{8}~\msun$ \citep[see
Fig.~\ref{fig:MW_KH},][]{bruns05, nidever08, nidever10}.

Finally, the mass scale at which stripping begins to be effective
should depend directly on the the properties of the host system
(i.e.~$n_{\rm halo}$ and $V_{\rm sat}$ in our analysis).
That is, these parameters should, on average, scale with the mass of
the host, such that a typical infalling satellite would experience a
stronger ram pressure in more massive host systems and the critical
mass scale for satellite quenching would increase accordingly.
While values of $n_{\rm halo}$ and $V_{\rm sat}$ consistent with those
expected for the Milky Way lead to an onset of stripping at roughly
$\sim10^{8-9}~\msun$, observations of more massive host systems, such
as rich groups and clusters, should yield high quenched fractions at
yet higher masses as the critical quenching mass increases with host
mass.
This picture is supported by observations of local clusters, which
find very few star-forming satellites at stellar masses of
$10^{9.5}~\msun$ versus the $\sim 30\%$ quenched fraction measured for
group and Milky Way-like hosts \citep[e.g.][]{smith12, boselli14b,
  phillips15a, sj16}.


\begin{figure}
 \centering
 \hspace*{-0.2in}
   \includegraphics[width=0.5\textwidth]{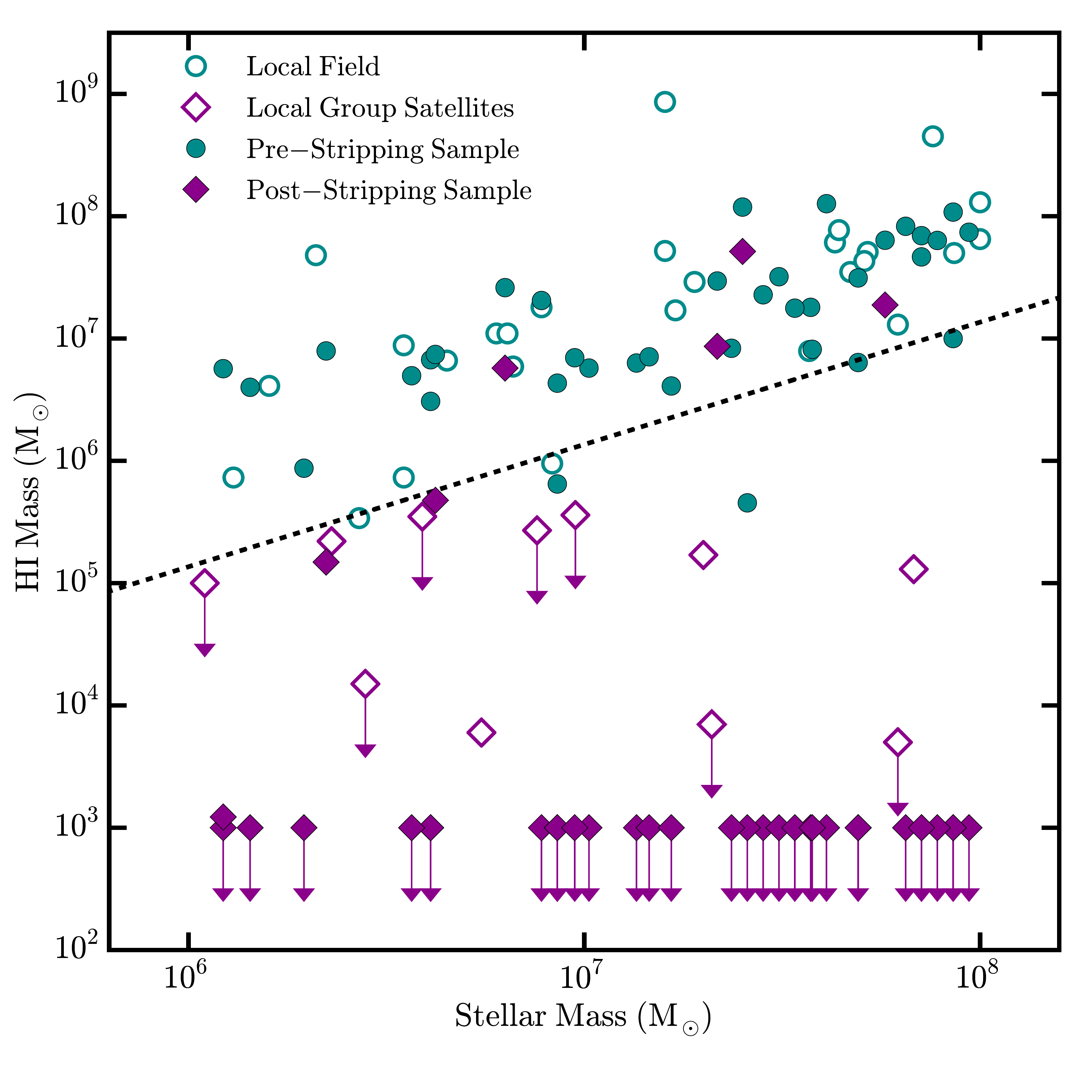}
   \caption{The H{\scriptsize I} gas mass as a function of stellar
     mass for our galaxy sample at $\mstar < 10^{8}~\msun$, prior to
     infall (solid cyan circles) and after interaction with the host
     CGM (solid magenta diamonds). To account for potential variations
     in the local halo gas density, we assume $n_{\rm halo} =
     10^{-3}$~cm$^{-3}$ and $V_{\rm sat} = 300$~km~s$^{-1}$, including
     both ram pressure and turbulent viscous effects in our stripping
     calculations. For comparison, we show the observed H{\scriptsize
       I} gas masses for field galaxies in the Local Volume (open cyan
     circles) and satellites of the Local Group (open magenta
     diamonds) from \citet{mcconnachie12} and
     \citet{spekkens14}. Arrows indicate observed upper-limits or
     systems that are completely stripped in our analysis. The dashed
     black line corresponds to an H{\scriptsize I} gas fraction of
     $0.136$, below which we define a galaxy as quenched. Including a
     clumpy host CGM, we find that stripping is able to quench
     $\sim90\%$ of our infalling satellite population at low masses
     ($33$ out of $37$ systems).}
 \label{fig:gasfrac}
\end{figure}


\subsection{Does Stripping Quench Low-Mass Satellites?}
\label{subsec:efficacy}

While ram-pressure stripping acts at the correct mass scales, our
analysis finds that ram pressure alone is unable to suppress star
formation in infalling satellites on the timescales predicted by
\citet{fham15}.
As shown in Fig.~\ref{fig:MW}, for expected values of $n_{\rm
  halo}$ and $V_{\rm sat}$, infalling satellites are typically
ram-pressure stripped of $<50\%$ of their cold gas reservoirs.
Given the typical H{\scriptsize I} gas fractions and star formation
rates for low-mass field dwarfs, which imply exceptionally long
depletion timescales \citep{skillman03, geha06, schiminovich10}, a
satellite stripped of only $50\%$ of its cold gas will still retain
enough fuel to potentially form stars for many Gyr. 
For $n_{\rm halo} = 10^{-3.5}$~cm$^{-3}$ and $V_{\rm sat} =
300$~km~s$^{-1}$, the gas fractions for our satellite population,
following ram-pressure stripping, are still typically $\sim3\times$
greater than the current observational limits for quenched satellites
in the Local Group \citep[i.e.~$f_{\rm H\scalebox{0.45}{\rm
    I}}\lesssim0.136$,][]{spekkens14}; 
that is, ram pressure only quenches $\lesssim15\%$ of our low-mass
satellite population, such that $f_{\rm H\scalebox{0.45}{\rm I}} < 0.136$.
Ultimately, to reproduce the observed H{\scriptsize I} gas fractions
for satellites in the Local Group and thus the inferred satellite
quenching timescales at low stellar masses, ram-pressure stripping
would need to be substantially more efficient than our predictions
(i.e.~stripping nearly the entire cold gas reservoir of all
systems, $<\!\!f_{\rm stripped}\!\!> \, \sim 0.9$).
\citet{emerick16} come to a similar conclusion based on wind-tunnel
modeling of an idealized Leo T-like satellite during infall.
Using the FLASH hydrodynamics code \citep{fryxell00}, they find that
ram-pressure stripping is unable to fully strip the satellite within
$2$ Gyr.

With the inclusion of viscous effects in our fiducial model, stripping
is able to remove the majority of cold gas from the low-mass,
infalling satellite population (i.e.~at $\mstar < 10^{8}~\msun$, see
Fig.~\ref{fig:MW_KH}).
In $60\%$ of low-mass systems following stripping, we find
H{\scriptsize I} gas fractions consistent with the observed limits for
the Milky Way dwarf spheroidal population ($f_{\rm
  H\scalebox{0.45}{\rm I}} < 0.136$), such that stripping is nearly
able to reproduce the high satellite quenched fraction observed in the
Local Group ($f_{\rm quenched} \sim 0.9-1$).
For roughly $40\%$ of systems, however, the resulting gas fractions --
post stripping -- are still greater than that observed for satellites
of the Milky Way.
Again, to bring our satellite population into agreement with the
roughly $90-100\%$ satellite quenched fraction at low masses in the
Local Group requires yet stronger stripping, such that the typical
stripped fraction is closer to $f_{\rm stripped} \sim 0.9$ at $\mstar
\lesssim 10^{8}~\msun$.

While current observations of comparable nearby systems (e.g.~M81,
M106) find low-mass satellite populations that roughly mirror that
found in the Local Group \citep{kaisin13, spencer14}, it remains
possible that the low-mass satellite quenched fractions for the Milky
Way and M31 are abnormally high relative to comparable host halos.
As shown in Fig.~3 of \citet{fham15}, if the satellite quenched
fraction for a Milky Way-like halo is typically $\sim70\%$ (versus
$90-100\%$), the satellite quenching timescale increases to
$\sim4-5$~Gyr (versus $\sim1-2$~Gyr). 
In such a scenario, stripping would not need to fully quench infalling
satellites; instead, stripping could simply remove roughly half of a
satellite's cold gas supply, so as to decrease the depletion
(i.e.~starvation) timescale accordingly.
Our analysis suggests that this level of stripping is very much
realistic for a Milky Way-like environment, even from ram pressure
alone.

\subsection{The Efficacy of Stripping: Refining Our Analysis}

Within our analysis, there are several factors or approximations by
which we are likely over- or under-estimating the true effectiveness
of stripping.
For example, we assume instantaneous ram-pressure stripping, which
likely overestimates the ram pressure \citep{tonnesen16}.
Moreover, like many studies of ram-pressure stripping, we adopt a
smooth host halo in equilibrium with the dark matter potential. X-ray
observations, however, find that massive hot halos, typically
associated with galaxy clusters, exhibit structure on scales of $\lesssim
1$~Mpc \citep{buote96, schuecker01}.
In addition, quasar absorption-line studies of low- and high-$z$ hosts
find significant clumpiness in the CGM of massive galaxies
\citep[e.g.][]{thom12, ab15}.
What the observed substructure in these systems implies for a typical
Milky Way-like object is unclear, but the assumption of a smooth halo
certainly ignores potentially important details.
Specifically, a clumpy gas halo will yield regions of higher density
and thus an increased ram pressure. Such a halo is also likely to have
some net velocity relative to the host's dark matter halo (our point
of reference in determining satellite velocities, see
Section~\ref{subsec:rhoV}). As shown by \citet{tonnesen08}, bulk
motion of the gas halo tends to be in the same direction as that of
the satellites, leading to a smaller effective $V_{\rm sat}$ and thus
weaker ram pressure.
Clearly, more detailed hydrodynamical simulations are needed to fully
address the impact of a clumpy CGM on our calculations.
However, to quench roughly $80-90\%$ of satellites in our sample via
stripping (i.e.~including ram-pressure and viscous effects) requires
local variations in the halo gas density on the order of $2-3\times$
that assumed in our fiducial model (i.e.~$n_{\rm halo} =
10^{-3.25}-10^{-3}$~cm$^{-3}$), given a satellite velocity of
$300$~km~s$^{-1}$. 
As shown in Figure~\ref{fig:gasfrac}, by including a clumpy CGM, our
analysis is able to reproduce the observed H{\scriptsize I} gas
fractions for the Local Group satellite population, with $\sim90\%$ of
our satellite population quenched following infall.
To achieve the same quenched efficiency with a satellite velocity of
only $200$~km~s$^{-1}$, thereby accounting for potential bulk motion
of the halo gas, our analysis requires a local increase in CGM density
of roughly $10-20\times$ that of our fiducial model (i.e.~$n_{\rm
  halo} = 10^{-2.5}-10^{-2.2}$~cm$^{-3}$).
While such extreme CGM densities are unrealistic on average, local
variations of this scale are in good agreement with the results of
recent hydrodynamic simulations of stripping by \citet{bahe15}, which
find that galaxies undergoing stripping in groups and clusters
typically experience increased ram pressure associated with CGM
overdensities as large as $100\times$ the mean.

Along with potential variations in the density of the host CGM, the
efficacy of stripping is also impacted by the density of the satellite
dark matter halo.
As discussed in Section~\ref{subsec:dMr}, the satellite dark matter
halo mass profiles assumed in our fiducial model yield a potentially
significant underestimate of the stripping efficiency.
If low-mass dwarfs live in less-concentrated host halos, as suggested
by their observed internal kinematics, the resulting stripped
fractions should increase by $\gtrsim 10\%$. 
With the inclusion of cored mass profiles, stripping becomes an
increasingly realistic physical driver for the high quenched fractions
and short quenching timescales for low-mass satellites.

While adopting less-concentrated mass profiles alone will likely not
be enough to fully strip all systems, it emphasizes another missing
ingredient in our analysis: baryonic feedback.
As shown by \citet{geha12}, field dwarfs (and thus infalling
satellites) are nearly universally star-forming at stellar masses of
$\lesssim10^{9}~\msun$ and will be subject to the stellar feedback
associated with radiation pressure and supernovae \citep[e.g][]{mqt05,
  hopkins14}.
As part of our analysis, we employ observed H{\scriptsize I} surface
density profiles for a sample of star-forming field dwarfs, along with
halo masses modeled from spatially-resolved kinematics, which should
thus naturally capture the impact of feedback on $\Sigma_{\rm gas}$
and $M(r)$.
Infalling satellites, however, may experience an elevated level of
star formation and thus feedback. 
This increased star-formation activity would result from compression
of the satellite's ISM due to interaction with the host CGM, thereby
increasing the local gas density and allowing star formation to
proceed at an accelerated rate.
For example, simulations of massive satellites in groups and clusters
find that ram pressure often leads to a burst of star formation for
the infalling system \citep{fujita99, bekki03, bekki14}.
Within the Local Group, modeling of the orbit and star-formation
history of Leo I suggest that it experienced a small burst of star
formation at pericentric passage prior to being quenched, consistent
with being initiated by ram pressure \citep{sohn13, bk13,
  weisz14a}.
This increase in star formation, driven by interaction with the host
CGM, can inject energy into the ISM of the satellite, puffing up the
system and thereby making it more susceptible to stripping
\citep[][but see also \citealt{emerick16}]{stinson07, el15}.
High-resolution hydrodynamic simulations of Milky Way- or Local
Group-like environments should provide a robust means for studying the
possible importance of feedback in increasing the efficiency of
stripping (e.g.~\citealt{zolotov12, mistani16, wetzel16};
Garrison-Kimmel et al.~in prep).


\begin{figure*}
 \centering
 \hspace*{-0.2in}
   \includegraphics[width=6in]{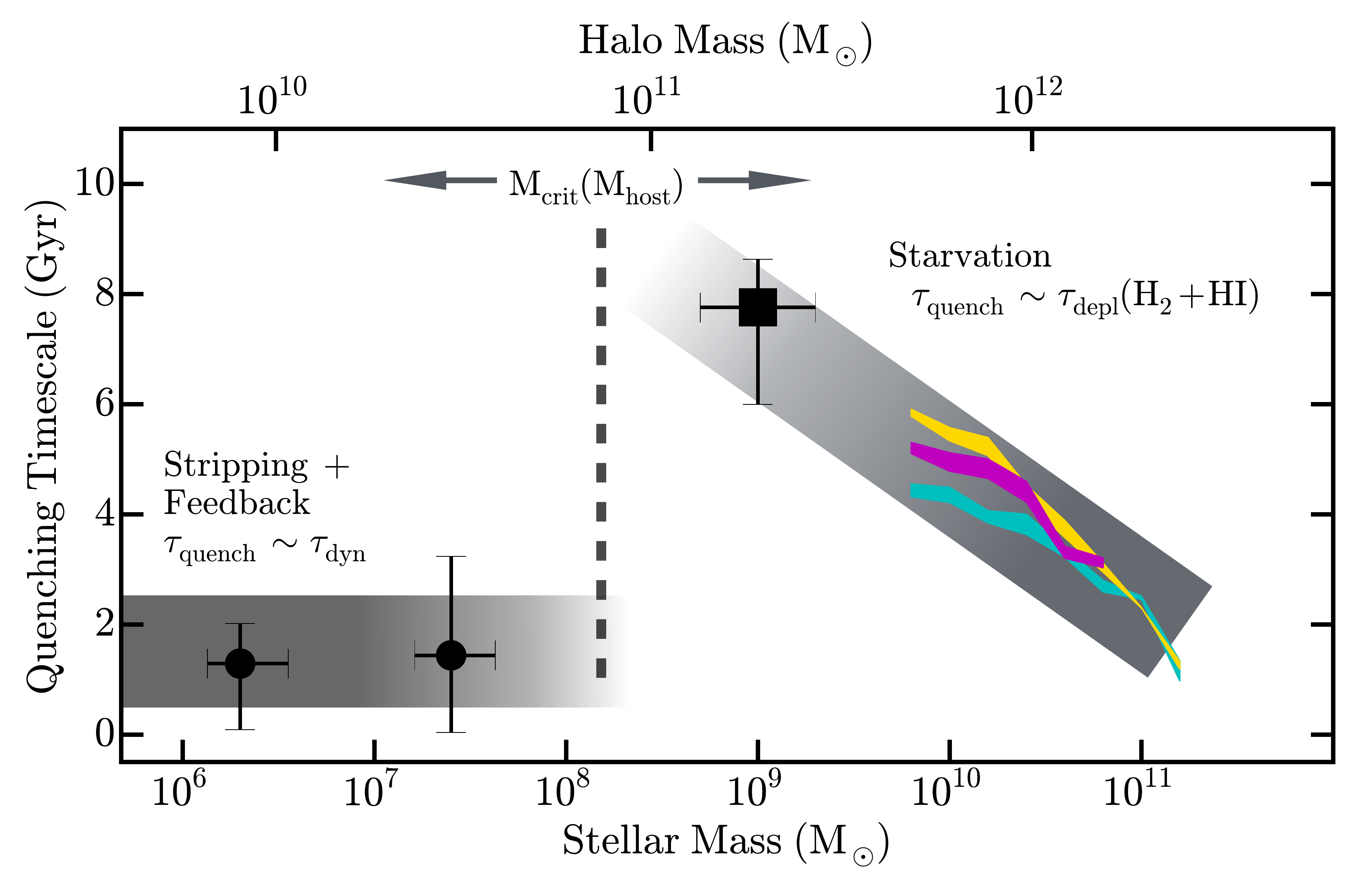}
   \caption{The dependence of the satellite quenching timescale on
     satellite stellar mass in Milky Way-like and more massive host
     halos ($>10^{12}~\msun$), as adapted from \citet{fham15}. The
     magenta, gold, and cyan colored bands show the constraints from
     \citet{wetzel13} for satellites in host halos of $M_{\rm host}
     \sim 10^{12-13}~\msun$, $10^{13-14}~\msun$, $10^{14-15}~\msun$,
     respectively. The black square and circles correspond to the
     typical quenching timescale for intermediate- and low-mass
     satellites from \citet{wheeler14} and \citet{fham15},
     respectively. The light grey shaded regions highlight the
     expected dominant quenching mechanism as a function of satellite
     mass, while the vertical dashed black line denotes the critical
     mass scale below which satellite quenching becomes increasingly
     efficient. At $\mstar \gtrsim10^{8}~\msun$, the satellite
     quenching timescales show broad agreement with the observed gas
     depletion timescales for field systems, suggesting that
     starvation is the main driver of satellite quenching at these
     masses. 
     At low masses, stripping -- potentially assisted by
     stellar feedback and a clumpy host CGM -- is the most probable
     mechanism responsible for the high satellite quenched fractions
     and short quenching timescales observed in the Local Group.
     The critical satellite stellar mass, $M_{\rm crit}$, at which the
     dominant quenching mechanism shifts from starvation to stripping
     should depend on the halo mass of the host system, with more
     massive hosts able to strip more massive satellites. }
 \label{fig:model}
\end{figure*}


\subsection{Towards a Complete Picture of Satellite Quenching}
\label{subsec:BigPic}

Figure~\ref{fig:model} presents current constraints on the satellite
quenching timescale from \citet{wetzel13}, \citet{wheeler14}, and
\citet{fham15} along with a qualitative depiction of the dominant
quenching mechanisms likely at play as a function of satellite stellar
mass.
As illustrated by \citet{fham15}, the measured satellite quenching
timescales at high masses, including the lack of strong dependence on
host halo mass, are broadly consistent with the expectations for
quenching via starvation \citep[see also][]{vdb08, wetzel13}.
At low masses, on the other hand, the short quenching timescales
inferred from analysis of the Local Group satellite population are
difficult to fully explain.
As our analysis shows, stripping is a likely culprit in suppressing
star formation at low masses, as it qualitatively reproduces the
critical mass scale for satellite quenching ($M_{\rm crit}$) within
the Local Group.
However, our analysis suggests that stripping (and specifically
ram-pressure stripping) may require a significantly clumpy host CGM or
the assistance of baryonic feedback to effectively remove enough cold
gas from the most gas-rich and concentrated systems.
Additionally, recent work by \citet{pearson16} and \citet{marasco16}
shows that the distribution of H{\scriptsize I} in low-mass field
galaxies can be significantly altered via close encounters with
neighboring dwarfs. 
If these encounters occur just before or during infall onto a host
system, the resulting satellite's ISM will likely be more susceptible to
stripping. 

If stripping drives quenching at low masses, then there are clear
implications regarding the CGM of the Milky Way and similar systems. 
In particular, to quench all low-mass satellites within $\sim2$~Gyr of
infall requires that the CGM extends to roughly $0.5~R_{\rm vir}$
\citep[or $\sim150$~kpc,][]{fham15} at a density of $n_{\rm halo}
\gtrsim 10^{-3.5}~{\rm cm}^{-3}$.
This large physical extent is needed to explain the quenching of
satellites with more circular or non-plunging orbits.
In ELVIS, we find that $\sim 25\%$ of subhalos in our selected mass
range and accreted at $0.15 < z_{\rm infall} < 3$ reach their first
pericentric passage at $0.5 < R/R_{\rm vir} < 1$.
To reproduce the extremely high satellite quenched fractions at
$\lesssim10^{8}~\msun$ via stripping, the CGM must therefore have a
relatively cored density profile \citep[e.g.][]{MB04}.

While the physical picture presented in Fig.~\ref{fig:model} broadly
explains the suppression of star formation in satellite systems, it
largely ignores any corresponding structural evolution. 
Recent observations at intermediate redshift suggest that quenching of
massive central galaxies is closely associated with development of a
bulge-dominated morphology \citep{bell12, cheung12}.
Moreover, observations of galaxy morphology in field and group/cluster
populations at $\mstar \gtrsim 10^{9.5}~\msun$ point towards an
evolution of satellite systems from disk- to bulge-dominated
\citep{vdb08, weinmann09}, suggesting that an additional mechanism
beyond starvation must likely be driving satellite evolution ---
unless fading of the stellar population, post quenching, can account
for observed differences in the light profiles of field and satellite
populations. 
In lower-mass hosts such as the Milky Way, however, there is little
morphological difference between massive satellites and field systems
of comparable mass \citep{phillips14}.
In addition, at lower satellite masses, structural evolution of
satellites is relatively modest following infall and is potentially
driven by baryonic feedback effects \citep{sj16, wheeler16}.

\section{Summary}
\label{sec:endgame}

Through the utilization of observed H{\scriptsize I} surface density
profiles for nearby dwarf galaxies, we investigated the effectiveness
of ram-pressure and turbulent viscous stripping in Milky Way-like
environments. 
Our analysis was motivated by recent results which point towards a
sharp change in the satellite quenching timescale, and therefore the
dominant quenching mechanism, for low-mass satellite galaxies. The
principal results of our analysis are as follows:

\begin{itemize}[leftmargin=0.25cm]

\item Ram-pressure and turbulent viscous stripping become increasingly
  effective in satellite galaxies with $\mstar~\lesssim~10^{9}~\msun$,
  consistent with the observed decrease in the satellite quenching
  timescale at $\mstar~\sim~10^{8}~\msun$. If stripping dominates the
  quenching of low-mass satellites, then we predict that the critical
  mass scale
  for satellite quenching should increase with host halo mass.  \\

\item Assuming a smooth host halo with a density of $n_{\rm halo} =
  10^{-3.5}~{\rm cm}^{-3}$ and a satellite velocity of $V_{\rm sat}
  \sim 300$~km~s$^{-1}$, we find that stripping is able to remove
  enough cold gas so as to quench $\sim60\%$ of infalling satellites
  at low masses.
  However, when including a clumpy halo, such that the typical CGM
  density at which stripping occurs is $n_{\rm halo} \sim 10^{-3.25} -
  10^{-2.5}~{\rm cm}^{-3}$ (i.e.~$\sim2-20$ times the mean density),
  stripping is able to effectively quench $\sim90\%$ of infalling
  satellites, such that their gas fractions agree with observational
  limits for dwarf spheroidals in the Local Group. \\

\item The efficiency of stripping may be further enhanced with the
  inclusion of stellar feedback, which could play an important role in
  making satellite systems susceptible to ram pressure and turbulent
  viscous effects. Further studies of stripping via hydrodynamic
  simulations will be a critical step in further constraining the role
  of stripping in the quenching of low-mass satellites.

\end{itemize}

\section*{acknowledgements}
We thank Stephanie Tonnesen, Adam Leroy, Fabian Walter, Manoj
Kaplinghat, David Buote, and David Jones for helpful discussions
regarding this work. We also thank Deidre Hunter and John Cannon for
providing H{\scriptsize I} data critical to the analysis.
Additionally, we thank the referee for providing helpful comments
which have improved the clarity of this work.
This work was supported in part by NSF grants AST-1518257,
AST-1517226, AST-1009973, and AST-1009999.
Support for ABP was provided by a GAANN fellowship. Support for SGK
was provided by NASA through Einstein Postdoctoral Fellowship grant
number PF5-160136 awarded by the Chandra X-ray Center, which is
operated by the Smithsonian Astrophysical Observatory for NASA under
contract NAS8-03060.
Support this work was provided by NASA through grants (AR-12836,
AR-13896 and AR-13888) from the Space Telescope Science Institute,
which is operated by the Association of Universities for Research in
Astronomy, Inc., under NASA contract NAS 5-26555.
MCC thanks the International Space Science Institute (ISSI) for
support of this work.

This research made use of {\texttt{Astropy}}, a community-developed
core Python package for Astronomy \citep{astropy13}. Additionally, the
Python packages {\texttt{NumPy}} \citep{numpy}, {\texttt{iPython}}
\citep{ipython}, {\texttt{SciPy}} \citep{scipy}, and
{\texttt{matplotlib}} \citep{matplotlib} were utilized for the
majority of our data analysis and presentation. Finally, we thank
Robert Van Winkle for his help in collaborating and listening.

\bibliography{rps}

\label{lastpage}
\end{document}